\documentclass[11pt ]{article}
\usepackage[latin1]{inputenc}
\usepackage[T1]{fontenc}
\usepackage[ english]{babel}
\usepackage[dvips]{graphicx}
\usepackage{amsmath}
\usepackage{palatino}

\title{Architecture of the Secoqc Quantum Key Distribution network}
\author{Mehrdad Dianati and Romain All\'eaume\\
GET-ENST, Network and Computer Science Department, CNRS UMR 5141\\
46 rue Barrault F-75634 Paris Cedex 13 France \\
email: $\{$dianati, alleaume$\}$@infres.enst.fr.
}

\begin{document}

\maketitle


\begin{abstract}
The European projet Secoqc (Secure Communication based on Quantum
Cryptography) \cite{Secoqc} aims at developing a global network
for unconditionally secure key distribution. This paper specifies
the requirements and presents the principles guiding the design of
this network, and relevant to its architecture and protocols.
\end{abstract}

\section{Introduction}

The performance of Quantum Key Distribution  (QKD) systems have
notably progressed since the early experimental demonstrations.
The current evolutions in QKD research \cite{Yam, Shields, Gisin,
Thew, HKLo} indicate that the pace of this progression is very
likely to be maintained, if not increased, in the future years. In
parallel to these improvements of QKD techniques, commercial
products are also being developed \cite{Magiq}, making QKD
deployment a feasible alternative for securing real data networks.

Deployment of a real QKD network is however far from being
straightforward. It requires development of a network architecture
connecting multiple users that may possibly be very far away from
each other. Considering the fact that the existing QKD links are
only point-to-point, and intrinsically limited in distance,
deployment of a practical QKD network structure is a nontrivial
problem, which is the target of European FP6 integrated project
Secoqc \cite{Secoqc}. As a part of the Secoqc project, in this
paper, we describe the proposed architecture for a QKD network. We
also specify the requirements relevant to the network design,
protocols, and services. The objective of this specification is to
define the major components and their main features.

The rest of this paper is organized as follows. In section
\ref{sec:qkd-overview} , we start by a brief account on the
interest and  applicability of quantum key distribution to secure
communications. In section \ref{sec:qkdnet}, we expose the
motivations for the development of QKD networks and provide a
survey of the previous works on QKD networks. Some major design
decisions of the Secoqc QKD network are discussed in this context
as well. In Section  \ref{sec:qkd-architecture}, we specify the
architectural view of the Secoqc QKD network, including the
components and their connections that are essential for the design
of network structure and protocols. Finally, in Section
\ref{sec:qkd-service-req}, we specify the services types provided
by the QKD network and discuss the important requirements related
to these services.

\section{Application of QKD to Secure Communications}
\label{sec:qkd-overview}


Distributing keys among a set of legitimate users while
guaranteeing the secrecy of these keys with respect to any
potential opponent is a central issue in cryptography, known as
the ``key distribution problem''. There are essentially two
methods currently in use to solve the key distribution problem
over deployed secure communication systems: 1) public key
cryptography and 2) secret couriers.

Public-key cryptography foundations rest on the difficulty of
solving some mathematical problems for which no polynomial
algorithms are known. The computing resources needed to solve
these problems  become totally unreachable when long enough keys
are used. Public-key cryptographic systems thus rely on what is
called computational security. Public-key cryptography is however
not unconditionally secure; the problems on which it is based are
not intractable; and in addition, their non-polynomial complexity
has so far not been proven.

The secret courier method is known since the ancient times: a
trusted courier travels between the different legitimate users to
distribute the secret keys, hopefully without being intercepted or
corrupted on his way by any potential opponent. Only practical
security can be invoked in this case, which has to be backed by
the enforcement of an appropriate set of security measures.
Although secret couriers become costly and unpractical when
implemented on large systems, this technique has remained in use
in some highly-sensitive environments  for which public-key
cryptography is thought to be inappropriate, such as government
intelligence or defense.
\\
Quantum Key Distribution, invented in 1984 by Charles Bennett and
Gilles Brassard \cite{BB84} based on some earlier ideas of Stephen
Wiesner \cite{Wiesner}, is an alternative solution for solving the
key distribution problem. In contrast to public-key cryptography
it has been proven to be unconditionally secure, i.e., secure
against any attack, whatever time, computing power or any other
resources that may be used  \cite{MayersProof, ShorPreskillProof}.
QKD security relies on the laws of quantum mechanics, and more
specifically on the fact that it is impossible to gain information
about non-orthogonal quantum states without perturbing  these
states \cite{Peres}. This property can be used to establish a
random  key between two users, commonly called Alice and Bob, and
guarantee that the key is perfectly secret to any third party
eavesdropping on the line, commonly called Eve.

\begin{figure}[!h]
\begin{center}
\includegraphics[width= 12cm]{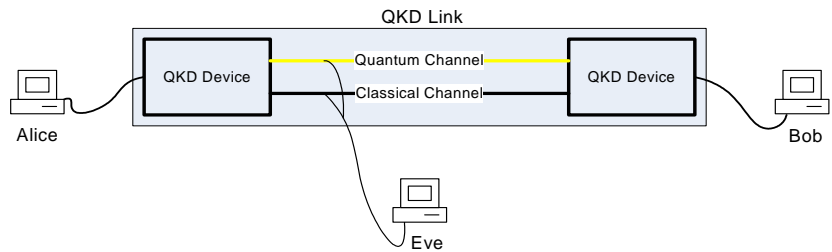}
\caption{Structure of a QKD link as it is referred throughout this
article} \label{fig:QKDLink}
\end{center}
\end{figure}

Without going in the details of the different implementations or
protocols, we can describe the structure and the principle of
operation of the basic practical QKD system: a QKD link. As
depicted on Fig. \ref{fig:QKDLink}, a QKD link is a
point-to-point connection between two users, commonly called Alice
and Bob, that want to share secret keys. The QKD link is
constituted by the combination of a quantum channel and a
classical channel. Alice generates a random stream of classical
bits and encodes them into a sequence of non-orthogonal quantum
states of light, sent over the quantum channel. Upon reception of
those quantum states, Bob performs some appropriate measurements
leading him to share some classical data correlated with Alice's
bit stream. The classical channel is then used to test these
correlations. If the correlations are high enough, this
statistically implies that no significant eavesdropping has taken
place on the quantum channel and thus that a perfectly secure
symmetric key can be distilled from the correlated data shared by
Alice and Bob. In the opposite case, the key generation process
has to be aborted and started again.

\section{QKD networks}\label{sec:qkdnet}

There are several fundamental limits regarding what can be
achieved with standalone QKD links. QKD links can by definition
only operate over point-to-point connections between two users,
which greatly restricts the domain of applicability of quantum key
distribution within secure communication networks. Furthermore,
since all QKD links rely on the transmission of quantum
information in order to guarantee security against on-line
eavesdropping, they will always remain intrinsically limited in
rate and distance, and cannot be deployed over any arbitrary
network topology. To overcome those limitations, it seems
important to study what can be achieved by networking QKD links in
order to extend the extremely high security standard offered by
QKD to the context of long distance communications between
multiple users. The development of QKD network architectures
appears from this perspective as a necessary step towards the
effective integration of QKD into secure data networks.

What we will call a QKD network throughout this paper is an
infrastructure composed of QKD links connecting multiple distant
nodes. The essential functionality of the QKD network is to
distribute unconditionally secure symmetric secret keys to any
pair of legitimate users accessing the network. These first
elements of definition are however fairly generic and can be
refined. Indeed, even though we are at the infancy of the
development of QKD networks, different models of QKD networks have
already been proposed. The first QKD network demonstrator, the
``DARPA Quantum network'', has been deployed between Harvard
University, Boston University and BBN in 2004 \cite{BBN, BBN2}.

It is convenient to characterize the different QKD network models
by the functionality that is implemented within the nodes. We can,
in this perspective, differentiate  three main categories of
network concepts, based on different ``families'' of node
functionalities :  1) optical ;  2)  quantum ; and  3) trusted
relay.

``Optical QKD nodes'' stands for nodes where  some classical
optical function, like beam splitting, switching, multiplexing,
demultiplexing, and etc., is implemented  on the quantum signals
sent over the quantum channel. The interest of such optical
networking functionalities is that they allow to go beyond 2-users
QKD. One-to-many connectivity between QKD devices was demonstrated
over a passively switched optical networks, using the random
splitting of single photons upon beam splitters \cite{Barnett}.
Active optical switching can also be used to allow the selective
connection of any two QKD nodes with a direct quantum channel. The
BBN Darpa quantum network \cite{BBN, BBN2} contains an active
2-by-2 optical switch in one node, that can be used to actively
switch between two network topologies. Optical functions can thus
be used to realize multi-user QKD and the corresponding nodes do
not need to be trusted, since quantum signals are transmitted over
an quantum channel with no interruption from one end user QKD
device to the other one. Such QKD network model can however not be
used to extend the distance over which keys can be distributed.
Indeed, the extra amount of optical losses introduced in the nodes
will in reality shorten the maximum span of quantum channels.

To be able to extend the distance on which key distribution can be
performed, it is necessary to fight against the degradation of
quantum signal along its propagation on the quantum channel. This
process, known as decoherence \cite{Zurek}, is typically quantum
and can only be taken care of by the use of  nodes which are able
to perform some active transformations on the quantum signals sent
over the quantum channel. We will call such nodes quantum nodes.
Quantum nodes can be of different flavors, but all rely on
entanglement, that can be used as a resource for secure key
generation \cite{EkertProtocol}. Indeed, as explained in
\cite{Biham}, a network of distant quantum memories each of them
able to store a share of a multipartite entangled states can be
used for quantum cryptographic purposes. In this perspective, the
main challenge of quantum nodes is to distribute entangled states
over long distances. The most elaborated quantum nodes proposed so
far are quantum repeaters \cite{Cirac}. They use entangled photon
sources, quantum memories and purification of entanglement to
obtain perfect entangled states, stored in nodes, over a quantum
channel segment. Such segments are then chained and entanglement
swapping used to obtain end-to-end perfect entanglement over an
arbitrary long distance. Quantum repeaters however rely on
elaborated quantum operations and on quantum memories that cannot
be realized with current technologies. As discussed in
\cite{Collins}, quantum nodes called quantum relays could also be
used to extend the reach of QKD. Quantum relays are simpler to
implement that quantum repeaters since the do they don't require
quantum memories. Building quantum relays remains however
technologically difficult and would not allow to extend QKD reach
to arbitrary long distances.

\begin{figure}[!h]
\begin{center}
\includegraphics[width= 12cm]{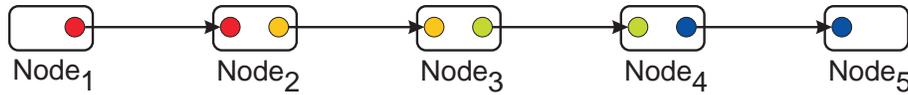}
\caption{``Hop-by-hop'' unconditionally secure message passing on
a path made of trusted relay nodes connected by QKD links. Message
decryption / re-encryption is done at each intermediate node, by
using one-time-pad between the local key, distributed by QKD,
$K_{local}$, and the secret message $M$ resulting in the ciphered
message $M \bigoplus K_{local}$. Different key association are
symbolized by different colors.  } \label{hop}
\end{center}
\end{figure}

Classical trusted relays can on the other hand be implemented with
today's technologies since such nodes consist in classical
memories, that we will call key stores, placed in secure
locations. QKD networks based on trusted key relays follow a
simple principle: local keys are generated over QKD links and then
stored in nodes that are placed on both ends of  each link. Global
key distribution is  performed over a QKD path, i.e. a
one-dimensional chain of trusted relays connected by QKD links,
establishing a connection between two end nodes, as shown on
Fig. \ref{hop}. Secret keys are forwarded, in a hop-by-hop
fashion, along QKD paths. To ensure their secrecy, one-time-pad
encryption and unconditionally secure authentication, both
realized with a local QKD key, are performed. End-to-end
information-theoretic security is thus obtained between the end
nodes, provided that the intermediate nodes can be trusted.
Classical trusted relays can be used to build a long-distance QKD
network. They have been used in the BBN QKD network \cite{BBN,
BBN2} and will also be used to build the Secoqc QKD network
\cite{Secoqc}.

The focus of the Secoqc project is on ``long-range high security
communications'' based on quantum key distribution''. As explained
above, this imposes to rely on trusted nodes used as key relays.
We have adopted this network model within the Secoqc project. An
important choice is however the way the local keys are used to
secure long-distance traffic. The main originality of the Secoqc
project with respect to previous works on QKD networks relies in
the fact that we have opted for a dedicated key distribution
network infrastructure that we have called ``network of secrets''
\cite{D-NET-03, DSEC17, DSYS10}. The functionality of the network of secrets is
solely to store, forward, and manage the secret key materials
generated by QKD. Such key distribution network is characterized
by dedicated link, network and transport layers and can be
considered somehow independently of the quantum key generation
processes and on the key requests arising from application.

The central design issue behind this concept is that the keys are
stored and managed within key stores, placed in nodes, and not
within QKD devices or within the machines running endpoint secure
applications. This design choice allows to manage keys over a
dedicated global network (the network of secrets) composed of key
stores linked together with classical channels. The network of
secrets is by essence a classical network but we will call it the
``Secoqc QKD network''  or ``the QKD network'' in the rest of this
paper since we will always assume that the underlying key
generation mechanism, responsible for filling the key stores, is
quantum key distribution. We present the architecture of this QKD
network in the next section.

\section{ Architecture of the Secoqc QKD Network}
\label{sec:qkd-architecture}

In this section, we explain the proposed architecture of the
Secoqc QKD network and its components. The purpose is to provide
an abstract perspective which gives a clear picture of the
components and the environment for designing the details of
network protocols. We also discuss the expected functional and
nonfunctional requirements of the QKD network.

\subsection{QKD Network Structure}

Many interesting network functionalities arise from the fact that
the network graph of QKD links and QKD nodes may provide redundant
paths between any two nodes. Within Secoqc, we have decided to
focus on such topologies for our QKD network. Such redundant
topologies are typically found in the core of communication
networks. This is the reason why we are referring to Quantum
Backbone (QBB) nodes for the central and multiply-connected nodes
within our QKD network.

\begin{figure}[!h]
\begin{center}
\centerline{\includegraphics[width=12cm]{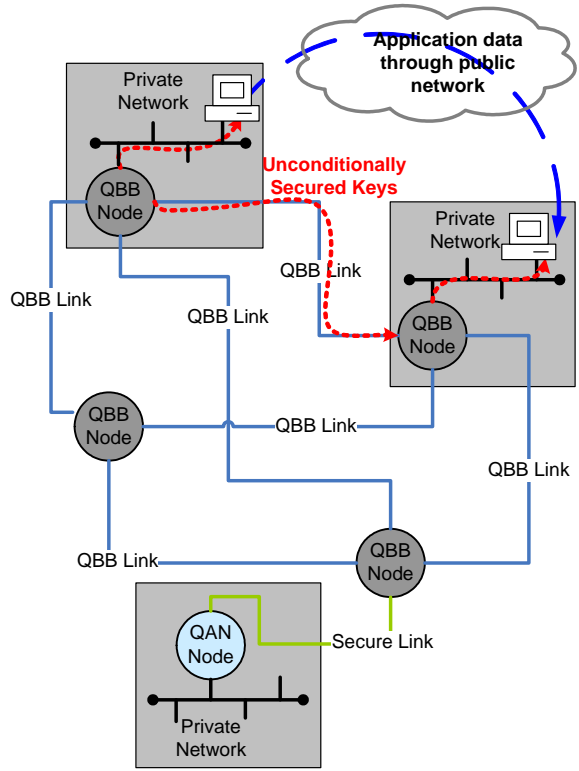}}
\caption{The structure of the Secoqc QKD network}
\label{fig:net-archintecture}
\end{center}
\end{figure}

The architecture of the Secoqc QKD network is illustrated in Fig. \ref{fig:net-archintecture}: the QKD
network is composed of a set of Quantum BackBone (QBB) nodes that
are connected by QBB links in order to distribute unconditionally
secure keys between any pair of application programs that are
running on host computers. The whole structure constitutes a
global key forwarding backbone network. In this network, the host
computers are connected to different QBB nodes across the network.
Alternatively, an application program can run on a host computer
which is connected to a Quantum Access Node (QAN). QAN is a node
with limited capabilities; it implements very little routing
functionalities, but is more specialized in providing access
interface to many client applications. As shown in Fig.
\ref{fig:net-archintecture}, the QAN is connected to a QBB node
via a secure connection, e.g., a QKD link. The application
programs that intend to have secure communications over a public
network, such as the Internet, can use the QKD network for
generation of unconditionally secure session keys. These session
keys,  can  then be used for cryptographic purposes. It has to be
emphasized that the QKD network will not carry the real data
traffic among application programs. The sole responsibility of the
QKD network is to share secret keys with unconditional security.
The actual communications, as shown in Fig.
\ref{fig:net-archintecture}, are performed on the existing public
or private networks.

Upon the request of an application program, namely the master
application program, the ingress QBB or QAN node, which is the
access point for the master application program, examines the
possibility of providing a path for the requested destination with
the desired Quality of Service (QoS). The destination application
program, namely the slave application program, obviously, must be
connected to another QBB or QAN node, which is referred to by the
egress QBB or QAN node. If the QKD network has enough resources,
it will accept the request for path establishment. Otherwise, the
request will be rejected. If the path is established successfully,
upon the request of the master application program, the ingress
QBB or QAN node generates a random session key and forwards it
through the QKD network to the egress QBB/QAN node, which will
deliver the key to the slave application program. In its way
through the QKD network the session keys will be secured through
hop-by-hop one-time-padding using the local keys. Local keys, that
are information theoretically secure, are established via QBB
links between adjacent QBB nodes. As shown in Fig.
\ref{fig:net-archintecture}, QBB nodes are located inside private
and secure networks (sites). Regarding that the local keys are
unconditionally secure and that the QBB nodes are inside secure
locations, we can conclude that the QKD network can be used for
global distribution of unconditionally secure session keys. Note
that the QBB links, which connect adjacent QBB nodes, are deployed
over potentially unsecured locations; thus, they are vulnerable to
different types of security threats. However, hop-by-hop
one-time-padding assures security of session keys. Furthermore,
redundant paths over the QKD network can be utilized by the
routing and forwarding protocols to combat Denial of Service (DoS)
attacks.

From the network point of view, QBB nodes are analogous to the
routers in the traditional inter-networking infrastructures. We
will discuss the main components of the network, i.e., the QBB
links and the QBB nodes, with further details in the subsequent
sections of this paper.

\subsection{QBB Links}

\begin{figure}[h!]
\begin{center}
\centerline{\includegraphics[width=12cm]{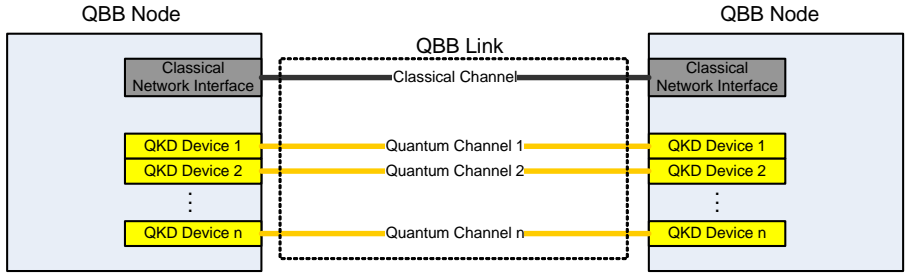}} \caption{
QBB link} \label{fig:qbb-link}
\end{center}
\end{figure}

The QBB links are special links, connecting QBB nodes together. As
shown in Fig. \ref{fig:qbb-link}, each QBB link consists of:
\begin{enumerate}

\item

arbitrary number of quantum channels to send quantum bits;

\item

a classical channel for signaling and forwarding of session keys.

\end{enumerate}

\begin{figure}[h!]
\begin{center}
\includegraphics[width= 12 cm]{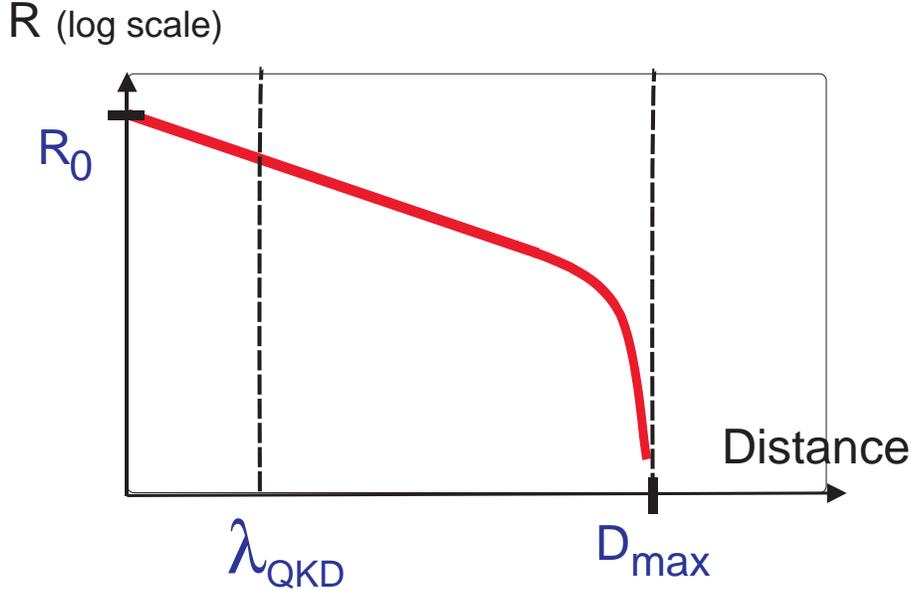}
\caption{Typical profile of the Rate versus Distance curve for a
single QKD link. } \label{fig:RateQKDLink}
\end{center}
\end{figure}

Quantum channels are used to send quantum bits, and the classical
channel is used for signalling and session key forwarding. There
might be several quantum channels between two adjacent QBB nodes
to improve the effective key generation rate. However, since an
extra pair of quantum devices is required to establish a single
quantum channel, the economical factors might hinder establishment
of multiple quantum channels. Referring to the classical notion of
QKD link in Section \ref{sec:qkd-overview}, a QBB link is an
extension of a QKD link, which allows to improve effective key
generation rate. The importance of this parallel structure can be
revealed if we consider the typical dependency between the length
of a QKD link and the effective key generation rate as shown in
Fig. \ref{fig:RateQKDLink}.
The vertical axis shows the effective key generation rate, and the
the horizontal axis shows the distance between the QKD devices.
Three essential parameters can be used to characterize the shape
of this curve:
\begin{enumerate}
\item

The secret bit rate at zero distance, $R_0$. It is typically in
the range 100 kbit/s to a few Mbit/s  for current technologies.

\item
 The scaling parameter $\lambda_{QKD}$.  As long as the signal to noise ration is high
enough and therefore the error rate well below some security
threshold, the logarithm of the secret bit rate scales linearly
with distance: $R(l)= R_0 \, e^{-l/\lambda_{QKD}}$. The scaling
parameter is platform and protocol dependent, but typically varies
between 5 and 25 km for QBB candidates platforms \cite{DQIT02}

\item

The maximum distance $D_{max}$. This distance is also platform and
protocol dependent and can vary from 20-30 km for short distance
systems to 100-120 km for long-distance systems.

\end{enumerate}

Stand-alone quantum channels are not sufficient to establish
unconditionally secure local keys. Further signalling through a
complementary classical channel is required to extract
unconditionally secure keys from the raw key materials, derived
from the measurements of quantum bits. In the Secoqc QKD network, the
session keys are forwarded by the QBB nodes in their path to the
destination over the classical channels. The classical channel is
also used for other purposes, e.g., routing messages, network
management messages, and etc. The classical channel is a virtual
channel rather than a physical channel. That is, it could be
either established over a public network through a TCP/IP socket
or it could be a direct point-to-point link between two adjacent
QBB nodes. The classical channels in the QKD network are not
secure channels. Proper cryptographic algorithms must be used if
the communication has to be secured using the local keys which are
generated by the QKD devices. For example, if some routing
messages have to be authenticated, a proper scheme must be
implemented.

It is important to note that quantum channels are subject to a
variety of security threats by adversaries. However, the
responsibility of QKD protocols \cite{BB84} is to generate local
keys between QBB nodes while being able to detect eavesdropping.
Two QKD devices will succeed to extract a secure key from the raw
key materials obtained from measurements of quantum bits provided
that the quantum bit error rate is below some security threshold
\cite{LutkenhausReview}. If the raw key bits contain too many
errors, security with respect to eavesdropping cannot be
guaranteed, thus, the raw key bits are discarded. Obviously, an
adversary can interrupt the key stream between two nodes by
measuring the quantum signals, introducing too many errors. One
objective of the QKD network is thus to enhance resiliency of key
distribution by providing redundant paths among QBB nodes. This
implies that the network protocols should be able to detect
possible attacks on the individual links and react to them by
choosing proper forwarding paths between QBB nodes.

\subsection{QBB Node}

QBB nodes are the main elements of the QKD network. Similar to
routers in a conventional network, they forward session keys
across the QBB network. They can also serve as access points for
key clients (i.e., application programs).
\begin{figure}[!h]
\begin{center}
\centerline{\includegraphics[width=12cm]{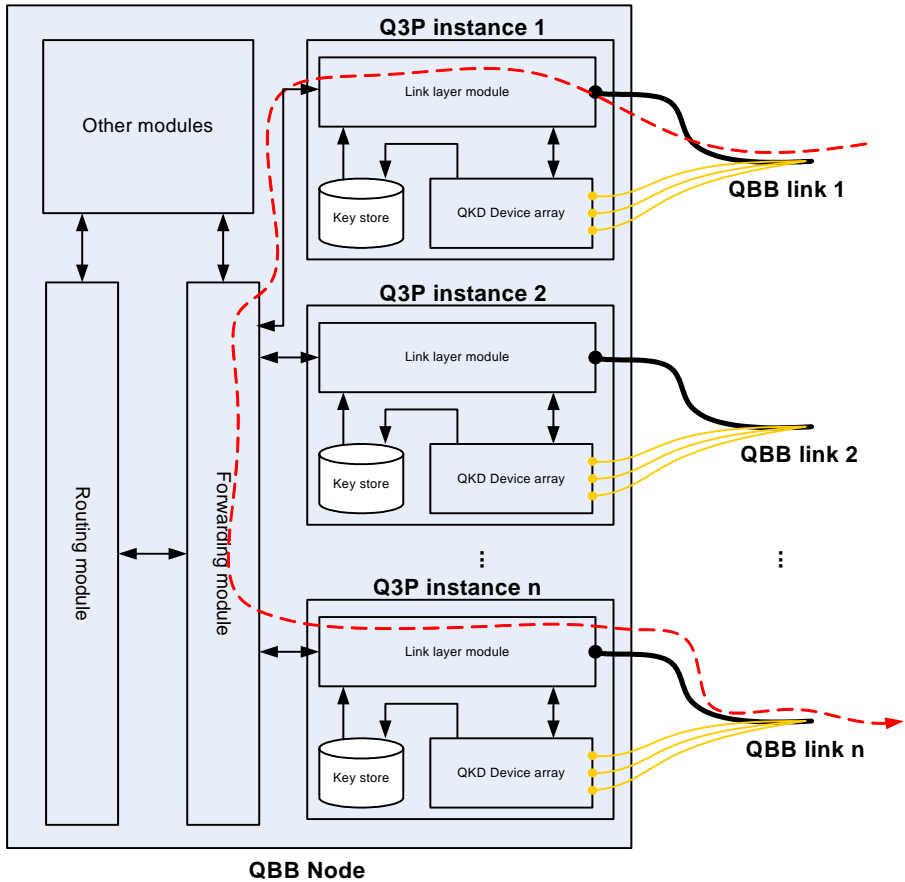}}
\caption{QBB node} \label{fig:qbb-node}
\end{center}
\end{figure}
Physically, QBB nodes are special computers with multiple QBB
interface ports, enabling connection to multiple QBB links. Each
QBB node can also have multiple QKD devices to create parallel QKD
links on each individual QBB link. The QKD devices can operate
over different types of quantum channels and rely on different
technologies, provided that they comply with the QBB node's
standard interface. There is also an interface for classical
channel on each QBB port. The logical structure of the software
components of a QBB node is shown in Fig. \ref{fig:qbb-node}. The
major software components are: 1) Multiple instances of Quantum
Point-to-Point Protocol (Q3P) \cite{D-NET-03} modules which serve
as the link layer modules for the QBB links; 2) a routing module
which is responsible for collecting and maintaining the local
routing information; and 3) a forwarding module for maintaining
paths and making forwarding decisions. All other software modules
such as node management, local connection management, random
session key generator, security monitor and etc. reside inside the
box denoted by "other modules". We do not elaborate on the other
modules in this paper as they have no or little interaction with
networking responsibilities of a QBB node.

Q3P module manages all link level communication issues between two
adjacent QBB node. A Q3P module includes a submodule for managing
the communications over a classical channel, including
functionalities for 1) authentication; 2) encryption/decryption;
3) fragmentation/assembling; 4) flow control; 5) link level error
control; 6) congestion control, if the classical channel is
established through a public network; 6) connection management
between two adjacent QBB nodes. Q3P module also includes a
submodule which acts as an array of device drivers for the QKD
device array. This module collects local keys from the QKD devices
and delivers them to a key store. Note that the QKD devices
constantly generate local keys at their maximum key generation
capacities. The QKD device drivers also allow the QKD devices to
share a single classical channel as it is needed for key
generation process inside the QKD devices. Another important
submodule of a Q3P module is the key store. The main purpose of
the key store is to compensate low key generation rate of QKD
devices by accumulating a big database of ready to use local keys
which will be used for local cryptographic operations between two
adjacent QBB nodes that are connected by a QBB link. Having the
key stores, QBB nodes can tolerate secret key traffic bursts by
buffering keys which are generated at a constant rate, but will be
consumed with a random rate.

To function as a router, the QBB node incorporates modules for key
forwarding and routing. The key forwarding module examines the
destination address or the virtual circuit number of the incoming
network key packets, performs necessary processing, and forwards
them to the appropriate output buffer corresponding to a
particular Q3P instance. An example of forwarding path inside a
QBB node is illustrated by the dashed curve in Fig.
\ref{fig:qbb-node}. A scheduling scheme prioritizes departure of
packets from the buffer to the destination Q3P instance according
to the QoS requirements of the buffered key packets. Forwarding
module also manages virtual circuits if necessary. This includes
call admission control, and traffic policing. Forwarding module
also creates and terminates virtual circuits inside a QBB node. In
order to enable key packet forwarding, a QBB node includes a
module for calculation and maintenance of routing information,
e.g., routing tables.

\section{Requirements of Services Provided by the Key Distribution Network
Layer}\label{sec:qkd-service-req}

In this section, we classify and specify the types of services
that will be provided by the Secoqc QKD network. This will help
selecting proper routing and forwarding scheme for the QKD
network. Simply put, the QKD network should provide
unconditionally secure and synchronized keys for the consuming
application peers. A typical scenario can be explained by
regarding the specified architecture in Section
\ref{sec:qkd-architecture}: an application program, running on a
host computer which is connected to a QBB or QAN node, intends to
open a secure connection to another application, running on
another host which is also connected to another QBB or QAN node.
An example could be an email client which is to setup an SSL
connection to a mail server. Another example is when a security
gateway attempts to establish a VPN connection through IPSec to
another security gateway. Currently, these applications rely on a
preset shared secret or certified public keys to generate session
keys. However, they can be modified to send their request to a
QBB/QAN node to establish a path through the QKD network to the
other end point. The key request should contain information about:
\begin{enumerate}

\item

the address of the destination node;

\item

the port number of the destination application;

\item

the QoS parameters such as key refreshment rate;

\item

the key block length.
\end{enumerate}
The ingress QBB node, which has received the request, tries to
establish such a connection through QKD network by examining the
local routing information and negotiation with the other QBB nodes
that might participate in establishing the connection. If the path
establishment is accomplished successfully, the ingress QBB node
replies to the request of the initiating application with a
positive message. In this case, the egress QBB node informs the
other party by sending an appropriate message. Otherwise, if the
path establishment fails due to insufficient available resources
or other reasons, appropriate acknowledgment is sent out to the
initiating application program. The failure acknowledgment message
could offer the initiating application with the best available
QoS. After path establishment process, the QKD network should
maintain the connection, and forward secret key which will be
generated by the ingress QBB node to the Egress QBB node reliably.

Regarding the aforementioned overview of the services of the QKD
network, some of the functional requirements of the QKD network
are:

\begin{enumerate}

\item

creation and maintenance of proper routing information which will
be distributed among QBB nodes;

\item

establishment and maintenance of paths between Ingress and Egress
QBB nodes;

\item

forwarding of secret keys by OTP through specified paths;

\item

coping with possible network and link level errors;

\item

monitoring security of the established path.

\end{enumerate}

In terms of QoS, the initial implementation of the QKD network
will provide the following service types.
\begin{enumerate}

\item

Best effort: for this type of service, a consuming application
will be allowed to transmit session keys with an average rate,
$\lambda_k$, and burst of $\sigma_k$. For this type of service,
the QKD network do not provide any guarantee of QoS; the
($\lambda_k$, $\sigma_k$) pair only specifies the upper bounds on
the request of the customer. In other words, a customer with this
type of service may receive less than what specified by the upper
bound. However, note that, regardless of the type of service, the
QKD network provides a reliable session key transport.

\item

Guaranteed key rate: for this type of service, the QKD network
guarantees a certain rate of session keys. More specifically, the
customer is allowed to ask for certain amount of keys within a
certain period of time. For instance, a customer may sign a
contract for 128 kbits of key materials every one second. Note
that in this context, the QKD network can only give a statistical
guarantee of service.

\end{enumerate}

An important nonfunctional requirement of QKD network is load
balancing by choosing proper paths which might not be the shortest
paths. This is important as the QBB links have limited capacities.
This also implies that the algorithms and schemes that heavily
rely on encrypted or authenticated signalling messages will not be
reasonable choices for the QKD network. Let's finally mention that
QKD network routing and forwarding algorithms also should be
resilient against link failure since such failures may be directly
due to the detection of security threats by some security
monitoring agents.

\section*{Acknowledgment}
We would like to thank the members of the NI and NET Secoqc
subprojects, and especially Louis Salvail, Oliver Maurhart, Alexander Marhold,
Thomas L\"anger, and Momtchil Peev for numerous helpful
discussions on the design of the network architecture. We
acknowledge financial support from the European Commission through
the IST-SECOQC Integrated Project.


\begin{thebibliography}{99}

\bibitem{Secoqc} www.secoqc.net
\bibitem{Yam} H. Takesue, E. Diamanti, T. Honjo, C. Langrock, M. M. Fejer, K. Inoue, Y. Yamamoto,{\it Differential phase shift quantum key distribution experiment over 105 km fibre}, quant-ph/0507110.
\bibitem{Shields} C. Gobby, ZL Yuan, and AJ Shields, {\it Quantum key distribution over 122km standard telecom fiber}, Appl. Phys. Lett. 84, 3762-3764(2004).
\bibitem{Gisin} N. Gisin, G. Ribordy, H. Zbinden, D. Stucki, N. Brunner, V. Scarani, {\it Towards practical and fast Quantum Cryptography}, quant-ph/0411022
\bibitem{Thew} R. T. Thew, S. Tanzilli, L. Krainer, S. C. Zeller, A. Rochas, I. Rech, S. Cova, H. Zbinden, N. Gisin, {\it  GHz QKD at telecom wavelengths using up-conversion detectors}, New J. Phys., Vol 8, 32 (2006),  eprint arxiv :quant-ph/0512054.
\bibitem{HKLo} Y. Zhao, B. Qi, X. Ma, H.-K. Lo, {\it Experimental Quantum Key Distribution with Decoy States}, Physical Review Letters 96, 070502 (2006),  eprint arxiv :quant-ph/0503192.
\bibitem{Magiq} www.magiqtech.com, www.idquantique.com




\bibitem{BB84}
C.H. Bennet, G. Brassard, "Quantum Cryptography: Public Key
Distribution and Coin Tossing", \emph{in Proc. of IEEE
International Conference on Computers, Systems, and Signal
Processing}, Bangalore, India, 1984, pp. 175-179.


\bibitem{Wiesner} S. Wiesner, \textit{Conjugate coding}, Sigact News, {\bf{15-1}}, 78-88 (1983). The original paper, written around 1970, had been refused for publication and remained unpublished until 1983.


\bibitem{MayersProof} D. Mayers, {\it Unconditionnal Security in Quantum Cryptography}, J. Assoc. Comput. Math. { \bf 48},  351, (1998), { \tt Eprint quant-ph/9802025}.


\bibitem{ShorPreskillProof} P. W. Shor et J. Preskill, { \it Simple Proof of Security of the BB84 Quantum
Key Distribution Protocol}, Phys. Rev. Lett., {\bf 85} (2000), 441Ð444 ; {\tt Eprint quant-ph/0003004}.


\bibitem{Peres}  A. Peres,{\it How to differentiate between non-orthogonal states},  Phys. Lett. A, vol. 128, pp. 19, Mar. 1988.





\bibitem{LutkenhausReview} M. Dusek, N. Lutkenhaus, M. Hendrych, {\it Quantum Cryptography}, eprint arxiv :quant-ph/0601207.




\bibitem{BBN}
C. Elliott, { \it Building the quantum network}, New J. Phys. {\bf 4} (July 2002) 46.

\bibitem{BBN2} Chip Elliott and al, {\it Current Status of The darpa Quantum Network}, eprint
arxiv :quant-ph/0503058, 2005.


\bibitem{Barnett}
 P. D. Townsend, S. J. D. Phoenix, K. J. Blow et S. M. Barnett, {\it Quantum cryptography for multi-user passive optical networks},  Electronics Letters, {\bf 30}, pp. 1875-1877 (1994).

\bibitem{Zurek} W. Zurek, {\it Decoherence and the transition from quantum to classical}, Physics Today, Volume 44, Issue 10, October 1991, pp.36-44

\bibitem{Biham} Eli Biham, Bruno Huttner, and Tal Mor, {\it Quantum Cryptographic Network based
on Quantum Memories}, Phys. Rev. A, 1996.
\bibitem{Cirac} J.Cirac, P.Zoller, and H.Briegel, {\it Quantum Repeaters based on Entaglement Purification}, eprint arxiv :quant-ph/9808065, 1998.


\bibitem{Collins} D.Collins, N. Gisin and H. de Riedmatten,  {\it Quantum Relays for Long Distance Quantum Cryptography},
eprint arxiv :quant-ph/0311101, 2003.


\bibitem{EkertProtocol} A. K. Ekert, { \it Quantum cryptography based on Bell's Theorem}, Phys. Rev. Lett. {\bf 67}, 661 (1991).

\bibitem{DSYS10} A. Marhold, D-SYS-10, {\it Technical Evaluation Report}, Secoqc Deliverable.

\bibitem{DSEC17} L. Salvail, Christian Schaffner, Requirements for security architectures (Rough network architecture for quantum communication applied to basic scenarios), Secoqc Deliverable D-SEC-17, Oct. 2004.

\bibitem{DQIT02} H. Bechmann-Pasquinucci, N. Cerf, M. Dusek, N. L\"utkenhaus, V. Scarani, M. Peev, {\it Report on a QIT-perspective comparison of the different platforms with respect to the evaluation criteria set in phase I of SECOQC}, Secoqc deliverable D-QIT-02, Sept. 2005.


\bibitem{D-NET-03} O. Maurhart, P. Bellot, M. Riguidel, R. All\'eaume, {\it  Network Protocols for the QKD network}, Secoqc deliverable D-NET-03, Oct. 2005.


\end{thebibliography}
\end{document}